\documentclass[%
 aip,
 amsmath,amssymb,
 reprint,%
]{revtex4-1}

\usepackage{graphicx}
\usepackage{dcolumn}
\usepackage{bm}

\usepackage[utf8]{inputenc}
\usepackage[T1]{fontenc}
\usepackage{mathptmx}
\usepackage{etoolbox}
\usepackage{graphicx,xcolor}
\usepackage{mathtools}
\usepackage{physics}
\usepackage{mathtools}
\usepackage{bbold}
\usepackage{amsthm}
\usepackage{tikz}
\usetikzlibrary{arrows,automata,positioning}
\usepackage{subcaption} 

\makeatletter
\def\@email#1#2{%
 \endgroup
 \patchcmd{\titleblock@produce}
  {\frontmatter@RRAPformat}
  {\frontmatter@RRAPformat{\produce@RRAP{*#1\href{mailto:#2}{#2}}}\frontmatter@RRAPformat}
  {}{}
}%
\makeatother

\newcommand{\ba}{\begin{eqnarray}}
\newcommand{\ea}{\end{eqnarray}}
\newcommand{\ban}{\begin{eqnarray*}}
\newcommand{\ean}{\end{eqnarray*}}

\newcommand{\one}{\mathbb{1}}

\newcommand{\aver}[1]{\left\langle #1 \right\rangle}

\newcommand{\valerio}[1]{{\color{black} #1}}

\newtheorem{result}{Result}

\newcommand{\xx}{x}
\newcommand{\yy}{y}
\newcommand{\dxx}{d_{\xx}}
\newcommand{\dyy}{d_{\yy}}
\renewcommand{\P}{\varphi}
\newcommand{\Ph}{\hat{\varphi}}
\newcommand{\Phg}{\Ph_{\gamma}}
\newcommand{\Phx}{\Ph_{\xi}}

\newcommand{\MPh}{M^{\hat{\varphi}}}
\newcommand{\MP}{M^{\varphi}}
\newcommand{\vxih}{v^{\hat{\xi}}}
\newcommand{\xih}{\hat{\xi}}
\newcommand{\vxi}{v^{\xi}}
\newcommand{\UU}{\mathcal{U}}
\newcommand{\EE}{\mathcal{E}}
\newcommand{\mSS}{\mathcal{S}}
\newcommand{\mUU}{\mathcal{U}}

\newcommand{\Pt}{\P(\yy|\xx)}

\newcommand{\bpm}{\begin{pmatrix}}
\newcommand{\epm}{\end{pmatrix}}

\begin{document}

\title[Fluctuation Theorems with Retrodiction rather than Reverse Processes]{Fluctuation Theorems with Retrodiction rather than Reverse Processes}

\author{Clive Cenxin Aw}
\affiliation{Centre for Quantum Technologies, National University of Singapore, 3 Science Drive 2, Singapore 117543, Singapore}

\author{Francesco Buscemi}
\affiliation{Graduate School of Informatics, Nagoya University, Chikusa-ku, 464-8601 Nagoya, Japan}

\author{Valerio Scarani}
\affiliation{Centre for Quantum Technologies, National University of Singapore, 3 Science Drive 2, Singapore 117543, Singapore}
\affiliation{Department of Physics, National University of Singapore, 2 Science Drive 3, Singapore 117542, Singapore}

\date{\today}

\begin{abstract}

Irreversibility is usually captured by a comparison between the process that happens and a corresponding ``reverse process''. In the last decades, this comparison has been extensively studied through fluctuation relations. Here we revisit fluctuation relations from the standpoint, suggested decades ago by Watanabe, that the comparison should involve the prediction and the retrodiction on the unique process, rather than two processes. We identify a necessary and sufficient condition for a retrodictive reading of a fluctuation relation. The retrodictive narrative also brings to the fore the possibility of deriving fluctuation relations based on various statistical divergences, and clarifies some of the traditional assumptions as arising from the choice of a reference prior.

\end{abstract}

\maketitle

\section{Processes versus inferences}

Quantitative approaches to irreversibility traditionally involve a comparison between the process physically happening, usually called forward process, and a corresponding reverse (or backward) process. The definition of the latter is intuitive only for some processes: somewhat ironically, not for those that are paradigmatic of irreversibility. Indeed, consider the erasure channel, which sends every possible input state to a unique, fixed output state: what should one take as its reverse process?

In a previous paper \cite{BS11}, two of us proposed to look at irreversibility as arising out of our logical inference, rather than out of physical processes. Specifically, we proposed to define the reverse process in terms of Bayesian retrodiction. This is a universal recipe. This retrodictive element can be identified \textit{a posteriori} in all previously reported fluctuation relations that we checked, including the most famous ones, both classical \cite{bochkov-kuzovlev-1977,jarzynski,crooks-theorem,jarz2000} and quantum \cite{tasaki2000jarzynski}, that are highlighted in the many available reviews \cite{campisi-haenggi-review-2011,jarzynski-review-2011,Seifert_2012,gawedzki2013fluctuation}. Besides recovering ``intuitive'' reverse processes, retrodiction provides a definition for the non-intuitive ones, which smoothly removes anomalies that were reported with other tentative definitions. Thus it seems plausible that all fluctuation relations can be understood in terms of retrodiction (though the literature is too vast and sparse to make a definitive call, we shall strengthen the evidence with Result \ref{res3} below).

In the pursuit of this line of research, we recognize that our previous paper was not radical enough. If the retrodictive origin of irreversibility is assumed, the narrative of the two processes becomes superfluous: using retrodiction to define a reverse process is an unnecessary step. \textit{There is only one process, the one that happens; what is being compared are our forward and backward inferences on it: prediction and retrodiction}. 

The replacement of irreversibility with \textit{irretrodictability} was pioneered by Watanabe \cite{watanabe55,watanabe65}, though prior to our previous work no connection had been drawn with the fluctuation theorems derived in the last twenty years. Under this change of viewpoint, it is the same physics that is being described, freed from an excess baggage in the narrative (and thus, possibly, on the interpretation). Besides epistemological economy, we are going to show that this viewpoint is \textit{fruitful} as it opens previously unnoticed vistas. 

The plan of the paper is as follows. In Section \ref{sec:retro} we present a self-contained introduction to retrodiction, both classical and quantum; and Section \ref{sec:examples} describes two case studies in detail. Section \ref{sec:ft} deals with \textit{fluctuation relations}: we show that these relations are intimately related with statistical distances (``divergences'') and that Bayesian retrodiction arises from the requirement that the fluctuating variable can be computed locally. We also compare the fluctuation relations obtained in the retrodictive narrative with those obtained in the reverse-process narrative. Section \ref{sec:priors} reflects back on the structure of retrodiction, elaborating on the role of the unavoidable \textit{reference prior}.

A word on the presentation. This paper covers topics from statistics, thermodynamics, and quantum information. We have tried to keep the presentation self-contained. We have also adopted a compact approach to references: besides those that prove specific results, we shall cite mostly reference books and reviews, and occasionally a few works that we consider clear and exemplary, useful as entry points for the reader, without any expectation of being exhaustive.

\section{Retrodiction: generalities}
\label{sec:retro}

In this paper, we consider processes with discrete alphabets. The input of the channel is denoted $x\in\{1,2,...,d_x\}$ and the output $y\in\{1,2,...,d_y\}$. We shall always have in mind $d_x=d_y=d$, keeping the notation different only when clarity demands it.

Also, throughout the paper, we assume that all probability vectors have strictly positive elements; and also all channels have only strictly positive entries (with the exception of the permutation channels studied in subsection \ref{ss:ham}). Arbitrarily small entries would be indistinguishable from an exact zero, certainly in practice, and perhaps also in principle depending on one's understanding of probabilities.

\subsection{Bayesian retrodiction on classical information}

As basic setting of retrodiction, consider the most elementary form of statistical inference: at the output of a known channel $\varphi(y|x)$, one observes the outcome $y=y^*$, and wants to infer something about the input $x$. In this paper, we focus on \textit{Bayesian retrodiction}, whose goal is to \textit{update one's belief} on the distribution of $x$. This requires a prior belief, the \textit{reference prior}, denoted $\xi(x)$. The total prior knowledge is therefore captured by the joint probability distribution $P_\xi(x,y)=\xi(x)\varphi(y|x)$; in particular, the prior knowledge about $y$ is $\hat{\xi}(y)= \sum_x\xi(x)\varphi(y|x)$. When the knowledge on $y$ is updated to $y=y^*$, one performs the \textit{Bayes' update}
\ba\label{bayes1}
P_\xi(x,y)&\stackrel{y=y^*}{\longrightarrow}&P_\xi'(x,y)=P_\xi(x|y^*)\delta_{y,y^*}
\ea
on the total knowledge, whence the updated knowledge on $x$ follows as $P_\xi(x|y^*)=\xi(x)\varphi(y^*|x)/\hat{\xi}(y^*)$. This is the most elementary example of retrodiction.

Slightly less basic, though also widely discussed in the statistical literature, is the retrodiction on $x$ based on ``soft evidence'' on $y$. This refers to the situation, in which the update on $y$ is not a sharp value $y=y^*$, but a distribution $u(y)$. In real life, soft evidence may arise by sheer uncertainty (e.g.~reading the outcome $y$ in very dim light) or by virtual evidence (e.g.~the doctor told me a definite result $z=z^*$ for my test, but I saw that he was tired and fear that he may have misread the actual result $y$ written on the sheet). The translation of such uncertainties into a quantitative $u(y)$ is not trivial \cite{CHAN200567,jacobs-changing-mind}, but we take it for granted. For such situations, Bayes' update \eqref{bayes1} is generalised to \textit{Jeffrey's update} \cite{jeffrey}
\ba\label{jeffrey}
P_\xi(x,y)&\,\stackrel{u(y)}{\longrightarrow}\,&P_\xi'(x,y)=P_\xi(x|y)\, u(y)\,.
\ea In the case of virtual evidence, Jeffrey's update is a direct consequence \cite{pearl,jaynes_2003} of Bayes' update starting from $z=z^*$, under the assumption that the variable $z$ influences directly only $y$ and not $x$ (c.f.~the example above: the tiredness of the doctor has no direct influence on whether I am actually sick). In other cases, it may be considered as an actual addition to the rules of Bayesian inference (this was Jeffrey's own view).

Thus, the conditional probability $P_\xi(x|y)$ plays the role of channel for the retrodiction, in short \textit{retrodiction channel}. For the remainder of the paper, we change the notation to
\ba\label{eq:phihat}
\hat{\varphi}_\xi(x|y)&=&\frac{\xi(x)}{\hat{\xi}(y)}\,\varphi(y|x)\,.
\ea

We shall make use at our convenience of a matrix representation. The channel $\Pt$ is represented by the column-stochastic matrix
\ban
\MP = \bpm \P(0|0) &  & \cdots &  & \P(0|\dxx) \\ & \ddots &  &  &  \\ \vdots &  & \Pt &  & \vdots  \\  &  &  & \ddots &  \\  \P(\dyy|0) &  & \cdots &  & \P(\dyy |\dxx ) \\  \epm
\ean
Similarly, the retrodiction channel $\hat{\varphi}_\xi(x|y)$ is represented by the column stochastic matrix
\ban
M^{\hat{\varphi}_\xi} = \bpm \Ph_\xi(0|0) &  & \cdots &  & \Ph_\xi(0|\dyy) \\ & \ddots &  &  &  \\ \vdots &  & \Ph_\xi(x|y) &  & \vdots  \\  &  &  & \ddots &  \\  \Ph_\xi(\dxx|0) &  & \cdots &  & \Ph_\xi(\dxx |\dyy ) \\  \epm\,.
\ean
In this notation, we similarly define inputs and outputs distributions $p(x)$ as column vectors $v^p$. For instance, the relations that define the reference prior can be written as
\ba\label{eq:Mxi}
\MP\vxi=\vxih&\;,\;& M^{\hat{\varphi}_\xi}\vxih= \vxi \,. 
\ea

\subsection{Two remarks}

Before continuing, we bring up two crucial remarks. The first is about the reference prior. It is well known \cite{watanabe55,watanabe65,jaynes_2003}, and our presentation above leaves it clear once again, that this element of subjectivity is an unavoidable feature of Bayesian retrodiction for a generic channel. The question of the \textit{choice of the prior} is a recurring topic in Bayesian statistics. The literature on fluctuation relations does not mention it as such, the assumption being stated in more physical language. We shall get back to this point in Section \ref{sec:priors}. Here, for the sake of definiteness we just mention two possible choices. One is the \textit{uniform prior} $\xi(x)=\frac{1}{d}$ for all $x$. Another is the \textit{steady state} of $\varphi$, defined by $\gamma=\hat{\gamma}$. Every stochastic map has at least one steady state, and exactly one if all its entries are strictly positive. It follows immediately from $\hat{\varphi}_\gamma(x|y)=\frac{\gamma(x)}{\gamma(y)}\,\varphi(y|x)$ that the uniform prior is a steady state if and only if $\varphi$ is bistochastic.

The second remark, that also others felt the need to highlight \cite{BJPsym21}, is that \textit{retrodiction is not inversion}. A channel has a linear inverse if there exists $M$ such that $M\MP=\one$. In the case of an invertible channel, given a valid output distribution $\hat{p}(y)=\sum_x\varphi(y|x)p(x)$, one is able to recover the input distribution $p(x)$. But for most invertible channels, $M$ is not a channel itself: there exist $u(y)$ such that $v^u\neq \MP v^p$. In particular, since the image of the probability simplex by $\MP$ is convex, there exists $y^*$ such that no input distribution $p(x)$ is mapped by $\MP$ to $\delta_{y,y^*}$ -- while retrodicting from $\delta_{y,y^*}$ is the most basic example of Bayesian inference. Ultimately of course the difference is in the task: retrodiction does not aim at reconstructing the prior through repeated sampling, but at updating one's belief after a single run of the process \footnote{This is particularly clear in the example of the test for a sickness given above. We can also refer the reader to the studies of retrodiction of quantum temporal dynamics \cite{GJM13,BJD+20}.}.

In subsection \ref{ss:ham}, we shall see a remarkable coincidence: the channels for which the retrodiction channel coincides with the inverse, and those for which the retrodiction channel is independent of the reference prior, are exactly the same.

\subsection{Retrodiction on ``quantum-inside'' classical channels}

\begin{figure}[h]
  \centering
  {\includegraphics[width=0.4\textwidth]{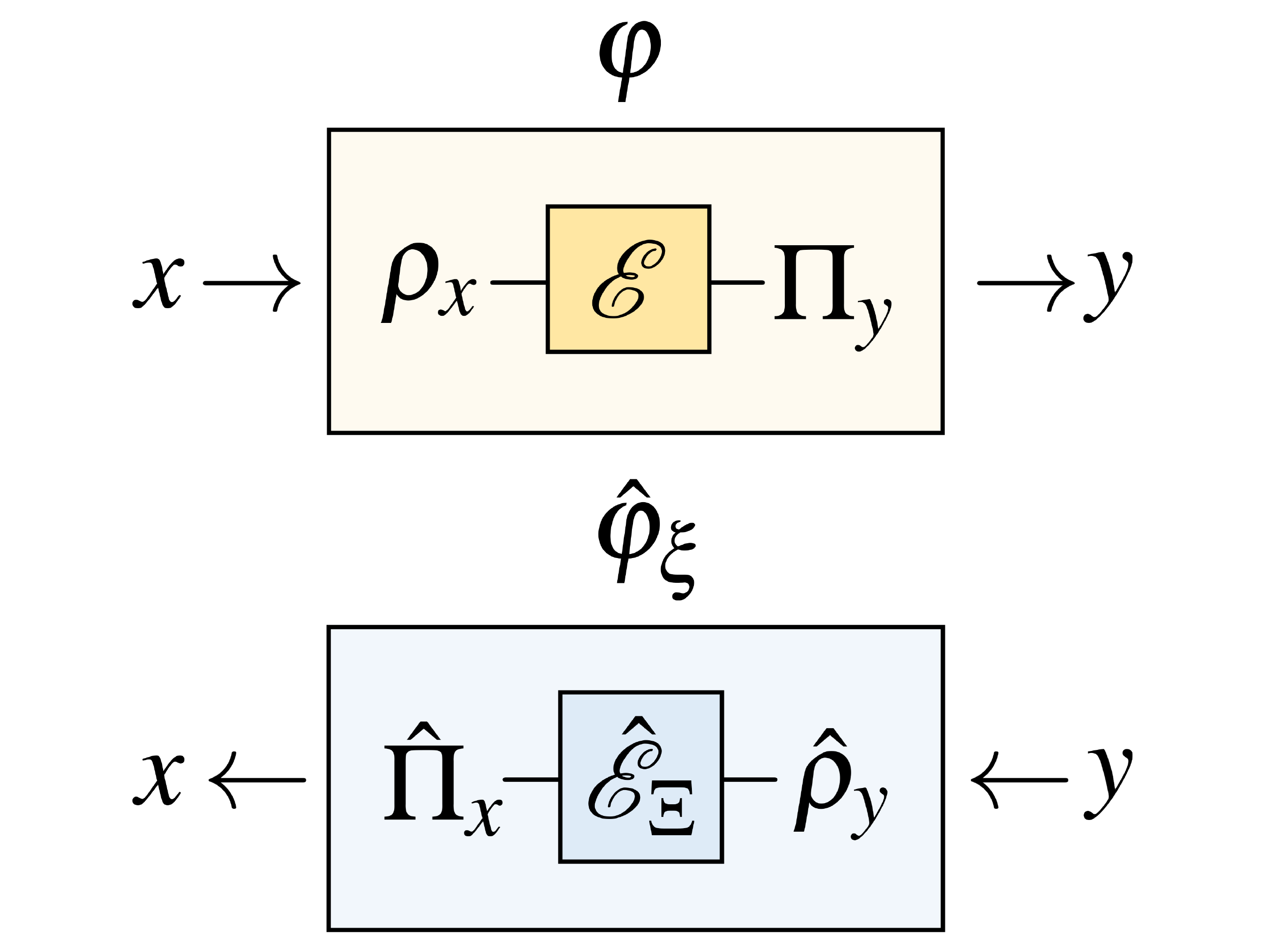}}
  \caption{A ``quantum-inside'' classical channel and the corresponding retrodiction channel. The construction, described in Eqs.~\eqref{statexi}-\eqref{eq:retropovm}, is valid for every set of states $\{\rho_x\}$, every CPTP map $\EE$ and every POVM $\{\Pi_y\}$.} 
  \label{fig:insideqm}
\end{figure}

According to our current knowledge, the most general description of the inner working of any classical input-output channel is given by quantum theory. The ``quantum-inside'' description of a classical channel is as follows (Fig.\ref{fig:insideqm}). The classical input $x$ prepares a system in a state $\rho_x$. The system is then sent through a quantum channel (a completely positive, trace-preserving, CPTP map) $\mathcal{E}$, and eventually measured with the positive operator-valued measure (POVM) $\{\Pi_y\}$, leading to the classical outcome $y$. All in all:
\ba\label{eq:qchannel}
\varphi(y|x)&=&\textrm{Tr}(\Pi_y\ \mathcal{E}[\rho_x])\,.
\ea
We want to derive the quantum description of the associated classical retrodiction channel \eqref{eq:phihat}: that is, finding states $\hat{\rho}_y$, a CPTP map $\hat{\mathcal{E}}$, and a POVM $\{\hat{\Pi}_x\}_x$, such that
\ba\label{eq:qretrochannel}
\varphi_\xi(x|y)&=&\textrm{Tr}(\hat{\Pi}_x\ \hat{\mathcal{E}}[\hat{\rho}_y])\,.
\ea 
For this, we first need to define the adjoint $\EE^\dagger$ of the channel, that is the operator such that $\textrm{Tr}(Y\EE[X])=\textrm{Tr}(\EE^\dagger[Y]X)$ for all operators $X,Y$. Inserting this definition in \eqref{eq:phihat}, one finds
\ban
\varphi_\xi(x|y)&=&\textrm{Tr}\left((\xi(x)\rho_x)\ \EE^\dagger[\Pi_y/\hat{\xi}(y)]\right)\,.
\ean This looks like \eqref{eq:phihat}, but in general one has $\textrm{Tr}(\Pi_y/\hat{\xi}(y))\neq 1$, $\sum_x \xi(x)\rho_x\neq \one$, and $\EE^\dagger$ is a CPTP map only if $\EE$ is unital. In order to identify proper states, channel and measurement, one has to introduce a \textit{reference state} 
\ba\label{statexi}
\Xi&=&\sum_x \xi(x)\rho_x\,.
\ea As in the classical case, we assume that $\Xi$ and $\hat{\Xi}=\EE[\Xi]$ have full rank, to skip caveats for situations of measure zero. Then one possible construction of the quantum elements of \eqref{eq:qretrochannel} uses
\ba
\hat{\rho}_y&=&\frac{1}{\hat{\xi}(y)}\mSS(\EE[\Xi])[\Pi_y]\,,\label{eq:retrostate}\\
\hat{\mathcal{E}}\equiv\hat{\EE}_\Xi&=&\mSS(\Xi)\circ\EE^\dagger\circ\mSS^{-1}(\EE[\Xi])\,,\label{eq:petz}\\
\hat{\Pi}_x&=&\xi(x)\mSS^{-1}(\Xi)[\rho_x]\,,\label{eq:retropovm}
\ea 
where we have introduced the notation $\mSS(A)[B]=\sqrt{A}\,B\,\sqrt{A}$ for a positive operator $A$. Starting from this basic constructions, one can obtain others:
\ba
\hat{\rho}_y&\longrightarrow&\UU_s[\hat{\rho}_y]\,,\\
\hat{\EE}_\Xi&\longrightarrow& \UU_m\circ \hat{\EE}_\Xi\circ \UU_s^{-1}\,,\label{eq:petzrot}\\
\hat{\Pi}_x&\longrightarrow&\hat{\Pi}_x\circ \UU_m^{-1}
\ea also lead to \eqref{eq:qretrochannel} for any pair of unitary channels $(\UU_s,\UU_m)$.

The key observation is that the retrodiction channel $\hat{\EE}$ turns out to be the \textit{Petz recovery} or \textit{Petz transpose} map \cite{petz1,petz} of $\EE$ for the reference state $\Xi$ [Eq.~\eqref{eq:petz}], or a rotated version thereof [Eq.~\eqref{eq:petzrot}].

The Petz map, a widely used tool in quantum information \cite{wilde_2013, SutterRennerFawzi_2016, AlhambraWoods_2017}, was previously identified on formal grounds as the generalisation of retrodiction within the quantum formalism \cite{fuchs2002quantum,Leifer-Spekkens,crooks-reversal,bergou-symm-retro}. First of all, in the case where all the states and the channels are diagonal in the same basis, \eqref{eq:petz} reduces to \eqref{eq:phihat}. Furthermore, just as the Bayesian retrodiction $\hat\varphi_\xi$ depends on a reference prior $\xi$, the Petz map $\hat{\EE}_\alpha$ depends on a reference state $\alpha$ \footnote{Petz discovered this map in the context of the monotonicity property of the quantum relative entropy $D(\rho||\alpha)=\Tr{\rho\ln\rho-\rho\ln\alpha}$~\cite{umegaki}. For any CPTP linear map $\EE$, one has $D(\rho||\alpha)-D(\EE[\rho]||\EE[\alpha])\geq 0$. Petz showed that $D(\rho||\alpha)=D(\EE[\rho]||\EE[\alpha])$ if and only if $\hat{\EE}_\alpha\circ\EE[\rho]=\rho$. In other words, $\hat{\EE}_\alpha$ reconstructs not only $\alpha$ after the channel, but every state whose relative entropy with respect to $\alpha$ has not changed.}. Interestingly, the Petz map was also used for quantum fluctuation relations \cite{kwon-kim}, but the connection with retrodiction was not noticed.

\section{Retrodiction: Two case studies}
\label{sec:examples}

In this Section, we present first retrodiction on Hamiltonian channels (both classical and quantum), which are provably the only ones for which the retrodictive map is independent of the reference prior and is identical to the inverse. Then we discuss retrodiction for all classical bit channels ($d=2$): precisely because it is elementary, this case study is useful to clarify features and dissipate possible confusions about retrodiction.

\subsection{Case study: Hamiltonian channels}
\label{ss:ham}

We call Hamiltonian channels, both classical and quantum, channels that are both deterministic and invertible (Watanabe \cite{watanabe65} referred to these channels as ``bilaterally deterministic''). The flows do not cross, and each state belongs to one and only one trajectory. 

For classical information, we have $y=f(x)$ with $f$ a bijection (in the discrete case, a permutation), and so $x=f^{-1}(y)$ is uniquely defined. In this case, it is absolutely natural to expect
\ba\label{eq:bijection}
\varphi(y|x)=\delta_{y,f(x)}&\,\Longrightarrow\,& \hat{\varphi}_{\xi}(x|y)=\delta_{x,f^{-1}(y)}
\ea independent of the reference prior. It is readily verified that this is indeed the case from Eq.~\eqref{eq:phihat}, since for a bijection we have $\hat{\xi}(y)=\xi(x)\delta_{y,f(x)}$.

This result has very appealing features: the retrodiction channel coincides with the inverse and is independent of the arbitrary choice of reference prior. Appealing as they are, these features cannot be taken as paradigmatic, because they are actually unique to this case.

\begin{result}\label{res1}
The following three statements are equivalent:
\begin{enumerate}
    \item[(\textbf{I})]  The channel $\P$ is a permutation.
    \item[(\textbf{II})] The retrodiction channel $\Ph$ is independent of the reference prior.
    \item[(\textbf{III})] There exists a reference prior $\xi$, for which the retrodiction channel $\Ph_\xi$ is the inverse channel $\P^{-1}$.
\end{enumerate}
\end{result}

\textit{Proof.} We present a full proof here, putting on record that the equivalence of (\textbf{{I}}) and (\textbf{{II}}) was already proved in Watanabe's pioneering study \cite{watanabe65}.

Eq.~\eqref{eq:bijection} proves (\textbf{{I}}) $\to$ (\textbf{{II}},\textbf{{III}}). The implication (\textbf{{II}}) $\to$ (\textbf{{III}}) goes as follows: Eq.~\eqref{eq:Mxi} implies
$M^{\hat{\varphi}_\xi}\MP\vxi=\vxi$. If $M^{\hat{\varphi}_\xi}=\MPh$ for all $\xi$, then $\MPh\MP\vxi=\vxi$ for all vectors: then  $M^{\hat{\varphi}}M^{\varphi}=\one$, that is $\hat{\varphi}=\varphi^{-1}$.

We are left to prove  (\textbf{{III}}) $\to$ (\textbf{{I}}). Let's assume that there is a reference prior such that $\hat{\varphi}_\xi=\varphi^{-1}$ i.e.~$M^{\hat{\varphi}_\xi}M^{\varphi}=\one$. Let us spell out this condition:
\ban
\quad M^{\hat{\varphi}_\xi}  \MP = \mathbb{1} & \Leftrightarrow & \quad\sum_y M^{\hat{\varphi}_\xi}_{x'y}  \MP_{yx} = \delta_{x'x}\quad\forall_{x'x} \\
& \Leftrightarrow & \quad   \sum_y \hat{\varphi}_\xi(x'|y) \varphi(y|x) = \delta_{x'x}\quad \forall_{x'x} \\
& \Leftrightarrow &  \quad  \sum_y \frac{\xi(x')}{\xih(y)}\varphi(y|x') \varphi(y|x) = \delta_{x'x}\quad \forall_{x'x}\,.
\ean
All the terms are products of non-negative numbers, $\xi(x)>0$ for all $x$ by assumption, and $1/\xih(y)\neq 0$ for $0\leq \xih(y)\leq 1$. Thus, for all the off-diagonal terms to be zero we need \ba\label{eq:offd}
\varphi(y|x') \varphi(y|x) = 0 \quad \forall_{x' \neq x, \, y}\,.
\ea 
This means the product of any two entries of a given $y$-row will always be zero. Hence, there can be at most one non-zero entry for that row, which means that the matrix $\MP$ can have at most $d$ non-zero entries. But there are $d$ columns, and the sum of all the elements of each column must be 1. Thus, the only possibility is that each row and each column have exactly one non-zero entry, and the value of the entry is 1. This defines a permutation matrix and concludes the proof. \qed

Incidentally, condition \eqref{eq:offd} shows that $M^{\hat{\varphi}_\xi}  \MP = \mathbb{1}$ is not determined by the reference prior; so at that point we had proved directly (\textbf{{III}}) $\to$ (\textbf{{II}}).\\

The same result holds for retrodiction on quantum information -- in fact, Result \ref{res1} was presented first for reasons of clarity, but can be seen as a special case of the following:

\begin{result}\label{res2}
The following three statements are equivalent:
\begin{enumerate}
    \item[(\textbf{I})]  The channel $\EE$ is unitary.
    \item[(\textbf{II})] The retrodiction channel $\hat{\EE}$ is independent of the reference prior.
    \item[(\textbf{III})] There exists a reference state $\alpha$, for which the retrodiction channel $\hat{\EE}_\alpha$ is the inverse channel $\EE^{-1}$.
\end{enumerate}
\end{result}

\textit{Proof.} The implications (\textbf{{I}}) $\to$ (\textbf{{II}},\textbf{{III}}) follow from the direct calculation of \eqref{eq:petz} for a unitary channel:
\ban
\hat{\UU}_\alpha&=&\mSS(\alpha)\circ\mUU^\dagger\circ\mSS^{-1}(\mUU[\alpha])\\
&=&(\mUU^\dagger\circ\mUU)\circ\mSS(\alpha)\circ\mUU^\dagger\circ\mSS^{-1}(\mUU[\alpha])\\
&=&\mUU^\dagger=\mUU^{-1}
\ean where we have used $\mUU^\dagger\circ\mUU=\mathcal{I}$ the identity channel, and $\mUU\circ\mSS(\alpha)\circ\mUU^\dagger=\mSS(\mUU[\alpha])$ that is $U\sqrt{\alpha}U^\dagger=\sqrt{U\alpha U^\dagger}$.

The proof of (\textbf{{II}}) $\to$ (\textbf{{III}}) is analog to that for classical information. Trivially, $\hat{\EE}_\alpha\circ\EE[\alpha]=\alpha$ holds by definition of $\hat{\EE}_\alpha$. Therefore, if $\hat{\EE}_\alpha=\hat{\EE}$ for all $\alpha$, then $\hat{\EE}[\EE[\rho]]=\rho$ for all $\rho$; whence $\hat{\EE}=\EE^{-1}$.

Finally, for the proof of (\textbf{{III}}) $\to$ (\textbf{{I}}): since any Petz map is CPTP, the starting assumption $\hat{\EE}_\alpha=\EE^{-1}$ implies that $\EE^{-1}$ is a CPTP map. But it is known that a CPTP map $\EE$ with the same input and output space has a CPTP inverse (that is, it is invertible, and the inverse is itself a channel) if and only if it is unitary~\cite{clean-povm}\footnote{If the input and output spaces are allowed to differ, a very similar results hold with an obvious generalisation~\cite{nayak-sen}: one has to append a system to match the dimensions before performing the unitary; and the inverse consists in undoing the unitary and tracing the additional system.}.
\qed

\subsection{Classical one-bit channels}

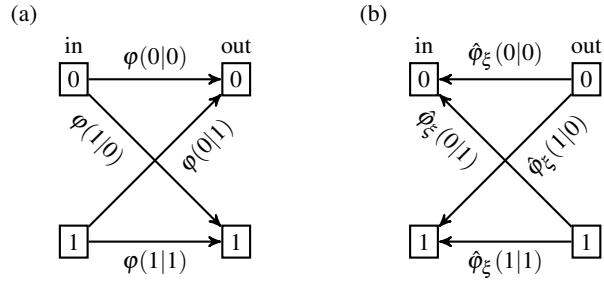
\begin{figure*}[t!]
\captionsetup[subfigure]{
        justification=raggedright,
        singlelinecheck=false,
        skip=0pt,
        margin=10pt
    }
    \centering
\begin{subfigure}[t]{0.25\textwidth}
    \caption{}
        \begin{tikzpicture}
            [scale= 0.9, >=stealth', auto, thick, diaglab/.style = {inner sep= 5pt, align=left, sloped}, nodey/.style = {shape=rectangle,draw}]
        \path 
        (-1.2,2.4) node (i0) [nodey, label=above:{in}] {0}
        (-1.2,0) node (i1) [nodey] {1}
        ( 1.2,2.4) node (o0) [nodey, label=above:{out}] {0}
        (1.2,0) node (o1) [nodey] {1};
        \draw[->] (i0) to node {$\P(0|0)$} (o0);
        \draw[->] (i1) edge node [diaglab, below, pos=0.75]{$\P(0|1)$} (o0);
        \draw[->]  (i0) edge node [diaglab, below, pos=0.2] {$\P(1|0)$}(o1);
        \draw[->] (i1) to node [swap] {$\P(1|1)$} (o1);
        \end{tikzpicture}
\end{subfigure}%
~
\begin{subfigure}[t]{0.25\textwidth}
    \caption{}
        \begin{tikzpicture}
            [scale= 0.9, >=stealth', auto, thick, diaglab/.style = {inner sep= 5pt, align=left, sloped}, nodey/.style = {shape=rectangle,draw}]
        \path 
        (-1.2,2.4) node (i0) [nodey, label=above:{in}] {0}
        (-1.2,0) node (i1) [nodey] {1}
        ( 1.2,2.4) node (o0) [nodey, label=above:{out}] {0}
        (1.2,0) node (o1) [nodey] {1};
        \draw[<-] (i0) to node {$\Phx(0|0)$} (o0);
        \draw[<-] (i1) edge node [diaglab, below, pos=0.75] {$\Phx(1|0)$} (o0);
        \draw[<-]  (i0) edge node [diaglab, below, pos=0.2] {$\Phx(0|1)$}(o1);
        \draw[<-] (i1) to node [swap] {$\Phx(1|1)$} (o1);
        \end{tikzpicture}
\end{subfigure}
    \caption{Bit channels (a) and their retrodiction (b) can be depicted as the respective maps above.}
    \label{fig:bitpictorial}
\end{figure*}

As a second case study, we consider classical stochastic processes for $d=2$ (Fig.~\ref{fig:bitpictorial}). We write a generic channel as
\ba\label{eq:bitchannel}
\MP = \bpm \P(0|0) & \P(0|1) \\ \P(1|0) & \P(1|1) \epm \,\equiv\,\bpm 1-a & b \\ a & 1-b\epm
\ea with $0\leq a,b\leq 1$. Its steady state is
\ban
\gamma(0)=\frac{b}{a+b}\,&,&\;\gamma(1)=\frac{a}{a+b}\,,
\ean unique unless $a=b=0$ (this being expected, since every state is a steady state for the identity channel).

The corresponding retrodiction channel with generic reference prior is 
\ba
M^{\hat{\varphi}_\xi} &=& \bpm \Phx(0|0) & \Phx(0|1) \\ \Phx(1|0) & \Phx(1|1) \epm \nonumber\\
&=& \bpm (1-a)\frac{\xi (0)}{\xih (0)} & a\frac{\xi (0)}{\xih (1)} \\ b\frac{\xi (1)}{\xih (0)} & (1-b)\frac{\xi (1)}{\xih (1)} \epm
\ea
with $\xih(0)\,=\,(1-a)\xi(0)+b\xi(1)$ and $\xih(1)=1-\xih(0)$. Interestingly, the retrodiction channel built on the steady state has the same stochastic matrix as the channel itself:
\ba\label{eq:RequalF}
M^{\Ph_\gamma}=\MP&\quad&[d=2]\,.\ea This can be verified without calculation, noticing that $\Phg(x|x)=\P(x|x)$ and that $M^{\Ph_\gamma}$ must also be column-stochastic.

The channel \eqref{eq:bitchannel} is \textit{invertible} if and only if $a+b\neq 1$. Result \ref{res1} of course holds: the retrodiction channel will be the inverse if and only if $a=b=0$ (identity channel) or $a=b=1$ (bit-flip channel). For all the other invertible channels, retrodiction and inversion do not coincide, whatever the choice of the reference prior.

The \textit{non-invertible channels}, $a+b=1$, make for an interesting case study; we change the notation $a\rightarrow \epsilon$, so that
\ba
\MP = \bpm 1-\epsilon & 1-\epsilon \\ \epsilon & \epsilon \epm \,.
\ea 
First notice that
\ba
\MP v^p=\bpm 1-\epsilon\\ \epsilon \epm
\ea for all input $p$. In other words, these channels erase whatever information is present in the input, and produce a fixed output distribution (which, of course, coincides with their steady state). In this sense, they could all be called \textit{erasure channels}, though the name is usually given to the case $\epsilon=0$.

\begin{figure*}[t!]
\captionsetup[subfigure]{
        justification=raggedright,
        singlelinecheck=false,
        skip=0pt,
        margin=10pt
    }
    \centering
\begin{subfigure}[t]{0.25\textwidth}
    \caption{}
    \begin{tikzpicture}
            [scale= 0.9, >=stealth', auto, thick, diaglab/.style = {inner sep= 5pt, align=left, sloped}, diaglabns/.style = {inner sep= 11pt, align=left}, nodey/.style = {shape=rectangle,draw}]
        \path 
        (-1.2,2.4) node (i0) [nodey, label=above:{in}] {0}
        (-1.2,0) node (i1) [nodey] {1}
        ( 1.2,2.4) node (o0) [nodey, label=above:{out}] {0}
        (1.2,0) node (o1) [nodey] {1};
        \draw[->] (i0) to node {$1-\epsilon$} (o0);
        \draw[->] (i1) edge node [diaglab, below, pos=0.75]{$1-\epsilon$} (o0);
        \draw[->, style=dotted]  (i0) edge node [diaglabns, below, pos=0.15] {$\epsilon$}(o1);
        \draw[->, style=dotted] (i1) to node [swap] {$\epsilon$} (o1);
        \end{tikzpicture}
\end{subfigure}
\begin{subfigure}[t]{0.25\textwidth}
    \caption{}
    \begin{tikzpicture}
            [scale= 0.9, >=stealth', auto, thick, diaglab/.style = {inner sep= 5pt, align=left, sloped}, diaglabns/.style = {inner sep= 11pt, align=left}, nodey/.style = {shape=rectangle,draw}]
        \path 
        (-1.2,2.4) node (i0) [nodey, label=above:{in}] {0}
        (-1.2,0) node (i1) [nodey] {1}
        ( 1.2,2.4) node (o0) [nodey, label=above:{out}] {0}
        (1.2,0) node (o1) [nodey] {1};
        \draw[<-] (i0) to node {$1-\epsilon$} (o0);
        \draw[<-, style=dotted] (i1) edge node [diaglabns, below, pos=0.9]{$\epsilon$} (o0);
        \draw[<-]  (i0) edge node [diaglab, below, pos=0.25] {$1-\epsilon$}(o1);
        \draw[<-, style=dotted] (i1) to node [swap] {$\epsilon$} (o1);
        \end{tikzpicture}
    \end{subfigure}
\begin{subfigure}[t]{0.25\textwidth}
    \caption{}
    \begin{tikzpicture}
            [scale= 0.9, >=stealth', auto, thick, diaglab/.style = {inner sep= 5pt, align=left, sloped}, diaglabns/.style = {inner sep= 11pt, align=left}, nodey/.style = {shape=rectangle,draw}]
        \path 
        (-1.2,2.4) node (i0) [nodey, label=above:{in}] {0}
        (-1.2,0) node (i1) [nodey] {1}
        ( 1.2,2.4) node (o0) [nodey, label=above:{out}] {0}
        (1.2,0) node (o1) [nodey] {1};
        \draw[<-] (i0) to node {$\xi(0)$} (o0);
        \draw[<-] (i1) edge node [diaglab, below, pos=0.8]{$\xi(1)$} (o0);
        \draw[<-]  (i0) edge node [diaglab, below, pos=0.25] {$\xi(0)$}(o1);
        \draw[<-] (i1) to node [swap] {$\xi(1)$} (o1);
        \end{tikzpicture}
\end{subfigure}
\caption{(a) An $\epsilon$-erasure channel, (b) the retrodiction channel with steady state as reference prior, (c) the retrodiction channel for a generic reference prior.}
    \label{fig:erasureall}
\end{figure*}
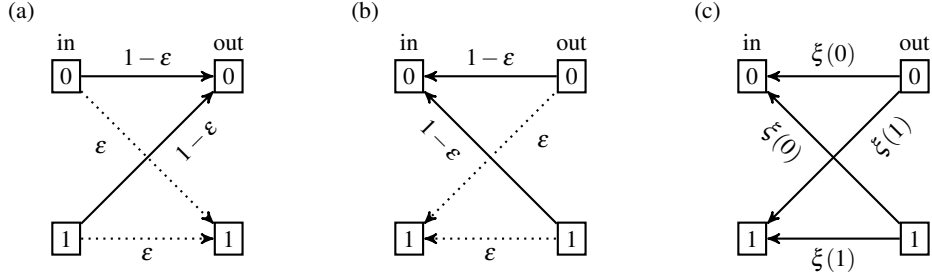

Because at the output all information on the input has been destroyed, one may naively expect the retrodiction channel to produce a completely random outcome. But this forgetting the importance of the reference prior in retrodiction. Plugging the expressions in the equations, one readily finds
\ba
M^{\hat{\varphi}_\xi} &=& \bpm \xi(0) & \xi(0) \\ \xi(1) & \xi(1) \epm \,.
\ea The retrodiction channel of an erasure channel is the erasure channel that returns the reference prior \valerio{-- a result that can be easily extended to any alphabet dimension \footnote{\valerio{Consider a generic erasure channel $\varphi(y|x)=\beta(y)$ for all $x,y\in [1,...,d]$,
that sends any input state to the state $\beta$. This will be the fate of any reference prior too: $\hat{\xi}(y)=\beta(y)$. Thus Eq.~\eqref{eq:phihat} yields
 $\hat{\varphi}(x|y)=\xi(x)$.}}.} In agreement with \eqref{eq:RequalF}, if the reference prior is the steady state, the retrodiction channel is the same erasure channel\footnote{The fact that one can associate a thermodynamically reversible process to a logically irreversible channel like the erasure channel was noted for instance by Sagawa \cite{Sagawa_2014}. The identity between the forward and the reverse erasure channels was noticed by Riechers and coworkers \cite{riechers2020} for a specific toy model of physical erasure. We see here that it is an unavoidable feature, having defined the reverse process from the steady state.}. These observations are summarized in Fig.~\ref{fig:erasureall}.

\section{Fluctuation relations from retrodiction}
\label{sec:ft}

\valerio{The topic of this section, fluctuation relations, originated in statistical thermodynamics. As we shall see, the formal structure of these relations can be derived without any reference to that branch of physics. As it happens, we shall mention thermodynamics only in the very last paragraph of the section. The explicit application of these formulas to important situations in thermodynamics was discussed in our previous paper \cite{BS11}.} 

\subsection{The process and its statistics}

As we noted in the introduction, it customary in studies of irreversibility to define the physical process as the \textit{forward process}, and to compare it to its corresponding \textit{reverse process}. Here we adopt a different narrative:
\begin{itemize}
    \item There is only one process, the one that is happening.
    \item A (forward) \textit{prediction} on the process starts with a prior $p(x)$ on the input, and infers the \textit{predicted distribution}
    \ba\label{defPF}
    P_F(x,y)&=&p(x)\varphi(y|x)\,.
    \ea
    \item A \textit{retrodiction} on the process starts with a prior $q(y)$ on the output, and infers the \textit{retrodicted distribution}
    \ba\label{defPR}
    P_R[\xi](x,y)&=&q(y)\hat{\varphi}_\xi(x|y)\,.
    \ea The explicit mention of $\xi$ will be dropped for simplicity in the remainder of this Section and resumed in Section \ref{sec:priors}.
\end{itemize}
We proceed to derive fluctuation relations with our narrative, and later we show the comparison with the reverse-process narrative.

\subsection{Derivation of the fluctuation relations}

Consider a variable $\Omega(x,y)$ that depends on the initial and final states, and may be determined by the process. Its predicted distribution is 
\ba\label{muF}
\mu_F(\omega)&=&\sum_{x,y}\delta(\omega-\Omega(x,y))\,P_F(x,y)\,,
\ea
while its retrodicted distribution is
\ba\label{muR}
\mu_{R}(\omega)&=&\sum_{x,y}\delta(\omega-\Omega(x,y))\,P_R(x,y)\nonumber\\
&=&\sum_{x,y}\delta(\omega-\Omega(x,y))\,R(x,y)\,P_F(x,y)
\ea
where
\ba\label{defrxy}
R(x,y)&=&\frac{P_R(x,y)}{P_F(x,y)}\,.
\ea
So, the difference between $\mu_F(\omega)$ and $\mu_R(\omega)$ is encoded this ratio of probabilities, which is exactly the quantity that appears in the \textit{statistical $f$-divergence} \cite{Csiszar-1967-f-div,liese-miescke}
\ba\label{fdiv}
D_f(P_R||P_F)&=&\sum_{x,y}P_F(x,y)\,f(R(x,y))
\ea
where the function $f(r)$ must be convex for $r\in\mathbb{R}^+$ and satisfy $f(1)=0$. The ``entropy production'', on which the thermodynamical literature bases fluctuation relations, uses $f(r)=-\ln(r)$, which generates the reverse Kullback-Leibler distance $D_{KL}(P_F||P_R)$. But we don't need to choose that particular function at this stage: for any function $f(r)$ invertible\footnote{Not all $f$ that are used for $f-$divergences are invertible. For instance, $f(r)=r\ln(r)$ for the Kullback-Leibler distance, or $f(r)=\frac{1}{2}|r-1|$ for the total variation distance. For such choices, one can still study the comparison between $\mu_R(\omega)$ and $\mu_R(\omega)$; it just won't take the compact form \eqref{ftretro}.} for $r\in\mathbb{R}^+$, if we set
\ba\label{assume}
\Omega(x,y)&=&f(R(x,y))
\ea we have by definition
\ba\label{avgfdiv}
\aver{\omega}_F&=&D_f(P_R||P_F)\,.
\ea
Besides, there immediately follows from \eqref{muF} and \eqref{muR} the \textit{fluctuation relation} \ba\label{ftretro}
\mu_{R}(\omega)&=&f^{-1}(\omega)\mu_F(\omega)
\ea that is the generalisation of Crooks' theorem \cite{crooks-theorem}. By integrating over $\omega$, one obtains the \textit{integral fluctuation relation}
\ba\label{ftintegr}
\left\langle f^{-1}(\omega)\right\rangle_F &=&\sum_{x,y}P_R(x,y)\,=\,1
\ea that depends only on the process. This is the generalisation of Jarzynski's equality \cite{jarzynski}.

\subsection{Comparison between retrodiction and reverse process}

In all the literature we are aware of, fluctuation relations are presented as a measure of the statistical difference \textit{between the forward and the reverse process, not between the predicted and retrodicted distributions of a single process}. The difference between the two narratives has mathematical manifestations that we are going to discuss now.

For the sake of definiteness, let us start with a canonical example. Suppose that the variable of interest is entropy, and that in the process under study it changes by $\Delta S$. In a retrodictive approach, \eqref{defPR} defines the retrodicted distribution for that same process. But if one looks at \eqref{defPR} as defining a reverse process, for that process the change of entropy will be rather $-\Delta S$.

Generalising this observation, the distribution of the variable $\omega$ in the reverse process reads
\ba
\tilde{\mu}_{R}(\omega)&=&\sum_{x,y}\delta(\omega-\tilde{\Omega}(x,y))\,P_R(x,y)\,,
\ea where, under assumption \eqref{assume},
\ba\label{defg}
\tilde{\Omega}(x,y)&=&f\left(\frac{1}{R(x,y)}\right)\,\equiv\,g(\Omega(x,y))
\ea because the roles of $P_F$ and $P_R$ are exchanged between the forward and the reverse process [for the choice $f(r)=-\ln(r)$, there follows the expected minus sign $g(\omega)=f(1/r)=-f(r)=-\omega$]. The resulting fluctuation relation then reads \cite{BS11}
\ba\label{ftreverse}
\tilde{\mu}_{R}(g(\omega))|g'(\omega)|&=&f^{-1}(\omega)\mu_F(\omega)\,.
\ea As expected, $\mu_F$ evaluated at $\omega$ is now related to $\tilde{\mu}_R$ evaluated at $g(\omega)$. The Jacobian factor, which comes from the change of variable in the $\delta$-function, ensures that the integral fluctuation relation takes exactly the same form as \eqref{ftintegr}.

A comparison of \eqref{ftretro} and \eqref{ftreverse} for various choices of $f$ is given in Table \ref{tab:fr}. For the thermodynamical case $f(r)=-\ln(r)$, we have $|g'(\omega)|=1$, and therefore the only difference between \eqref{ftretro} and \eqref{ftreverse} is that $\mu_R$ is evaluated at $\omega$ while $\tilde{\mu}_R$ is evaluated at $-\omega$. Thus in thermodynamics not only the Jarzynski equality, but also the Crooks fluctuation theorem is the same in both narratives (up to that sign change). Interestingly, even when reporting experiments in which the reverse process was actually implemented, it is the retrodictive version that is usually plotted for its visual convenience: see for instance the pioneering verification of Crooks' fluctuation theorem with folding and unfolding of RNA \cite{Collin05}.

\begin{table*}[t]
    \begin{tabular}{|c|l|c|c|c|}
      $\omega =f(r)$ & $f$-divergence & FR for retrodiction \eqref{ftretro} & FR for reverse process \eqref{ftreverse} & Integral FR \eqref{ftintegr} \\\hline\hline
      $-\ln(r)$ & Reverse Kullback-Leibler& ${\mu}_{R}(\omega)=e^{-\omega}\mu_F(\omega)$ & $\tilde{\mu}_{R}(-\omega)=e^{-\omega}\mu_F(\omega)$ & $\aver{e^{-\omega}}_F=1$\\\hline
      $(1-\sqrt{r})^2$& Squared Hellinger & $\mu_R(\omega)=(1-\sqrt{\omega})^2\mu_F(\omega)$ & $\tilde{\mu}_R\left(\frac{\omega}{(1-\sqrt{\omega})^2}\right)\frac{1+\sqrt{\omega}}{(1-\sqrt{\omega})^2}=(1-\sqrt{\omega})^2\mu_F(\omega)$ & $\aver{(1-\sqrt{\omega})^2}_F=1$ \\\hline
      $1/r-1$& Neyman $\chi^2$ & $\mu_R(\omega)=\frac{1}{1+\omega}\mu_F(\omega)$ & $\tilde{\mu}_R\left(\frac{\omega}{1+\omega}\right)\frac{1}{(1+{\omega})^2}=\frac{1}{1+\omega}\mu_F(\omega)$ & $\aver{\frac{1}{1+\omega}}_F=1$\\\hline
    \end{tabular}
    \caption{Fluctuation relations obtained in the retrodictive and in the reverse-process narratives, for a few choices of $\omega$ satisfying \eqref{avgfdiv} for the corresponding $f-$divergence. The fourth column is kept in the form \eqref{ftreverse} without possible algebraic simplifications, to facilitate the identification of $g(\omega)$ and $|g'(\omega)|$.}
    \label{tab:fr}
\end{table*}

\subsection{Fluctuation relations and Bayesian retrodiction}

In the retrodictive narrative, the fluctuation relation \eqref{ftretro} and its derivate \eqref{ftintegr} are statistical properties of the random variable $\omega$ defined by \eqref{avgfdiv}. They are formally valid for the statistical comparison between arbitrary $P_F$ and $P_R$, with no reference to the notion of retrodiction, let alone to its mathematical expression \eqref{eq:phihat}. In the reverse process narratives, one studies the distribution of the values of the variable when the roles of $P_F$ and $P_R$ are swapped [Eq.~\eqref{defg}]; but even then, the fluctuation relations follow without having specified any mathematical relation between $P_F$ and $P_R$. So, what is the role of Bayesian retrodiction, or that of a proper definition of the reverse process? We are going to prove that it singles out a specific structure for $R(x,y)$, and that this simple result has far-reaching consequences in the context of thermodynamics.

\begin{result}\label{res3}
The ratio $R(x,y)$ [Eq.~\eqref{defrxy}] is of the form $F(x)G(y)$, for some functions $F$ and $G$, if and only if $P_F$ and $P_R$ are related as \eqref{defPF} and \eqref{defPR}, with the latter constructed from Bayesian retrodiction [Eq.~\eqref{eq:phihat}]. In this case,
\ba\label{rxybayes}
R(x,y)&=&\frac{\xi(x)}{p(x)}\frac{q(y)}{\hat{\xi}(y)}\,.
\ea
\end{result}

\textit{Proof.} If $P_F$ and $P_R$ are given by \eqref{defPF} and \eqref{defPR}, using \eqref{eq:phihat} it is trivial to derive \eqref{rxybayes}. In the other direction: without loss of generality we keep the form \eqref{defPF} for $P_F$ and, using the product rule of joint probabilities, we write $P_R(x,y)=q(y)\eta(x|y)$ for the conditional distribution (channel) $\eta$ and marginal $q$. The assumption reads
\ban
\frac{\eta(x|y)}{\varphi(y|x)}\,=\,\tilde{F}(x)\tilde{G}(y)&\quad&\forall_{x,y}
\ean with $\tilde{F}(x)=p(x)F(x)$ and $\tilde{G}(y)=G(y)/q(y)$. Since the l.h.s.~is strictly positive\footnote{Recall our assumption that all the entries of the channel are positive; they could be as small as desired.}, $\textrm{sign}(\tilde{F}(x))=\textrm{sign}(\tilde{G}(y))$ must hold for all $(x,y)$. Now, being a channel, $\eta$ must satisfy $\sum_x\eta(x|y)=1$, that is $1/\tilde{G}(y)=\sum_x\tilde{F}(x)\varphi(y|x)$. So finally
\ban
\eta(x|y)&=&\frac{\tilde{F}(x)}{\sum_x\tilde{F}(x)\varphi(y|x)}\varphi(y|x)\,\equiv\,\hat{\varphi}_\xi(x|y)
\ean where $\xi(x)=\tilde{F}(x)/\sum_{x}\tilde{F}(x)$ is a valid probability distribution because all the $\tilde{F}(x)$ have the same sign. \qed

While this result may look purely anecdotal or formal, let us recall that in the usual thermodynamical interpretation
$\Omega(x,y)=-\ln(R(x,y))$ is the (non-adiabatic) stochastic entropy production~\cite{esposito-three-fluct}. Thus, \textit{whenever the stochastic entropy production can be computed locally (that is, independently of the \textit{correlations} between microstates $x$ and $y$), a structure of Bayesian retrodiction is unavoidable} (in the reverse process narrative: the reverse process must be defined through Bayesian retrodiction).

\section{On the choice of the reference prior}
\label{sec:priors}

In the literature on fluctuation relations, based on thermodynamics, the wording ``reference prior'' is absent. Its role is usually taken by an assumption of ``detailed-balance''. In all the examples that we have looked into \cite{BS11}, this corresponds to the choice of the \textit{steady state} as reference prior. The operational interpretation of this choice is very physical: one takes as reference the process in which nothing changes. It has also a very neat consequence when it comes to fluctuation relations: the ratio $R(x,y)$ given in \eqref{rxybayes}, and thus the variable that enters the fluctuation relations, depends on the channel $\varphi$ \textit{only} through its steady state $\gamma$. With this choice, one is clearly studying fluctuations \textit{around equilibrium}\footnote{We used the wording ``steady state'' instead of ``equilibrium'' because the latter has become almost synonymous of \textit{thermal} equilibrium. Our channels have nothing thermal \textit{a priori}, and even for channels involving contact with a thermal bath, the thermal state may not be the steady state (e.g.~if the Hamiltonian changes during the process).}

Inspired by statistical comparisons, one may opt for a different definition of the reference prior. One possibility is trying to keep prediction and retrodiction as close as possible. With such goals, let us take as a figure of merit
\ban
D_{KL}(P_F||P_R[\xi])&=&\sum_{x,y}P_F(x,y)\ln\left(\frac{P_F(x,y)}{P_R[\xi](x,y)}\right)\\&=&\sum_{x,y}P_F(x,y)\ln\frac{p(x)}{q(y)}+\sum_{x,y}P_F(x,y)\ln\frac{\hat{\xi}(y)}{\xi(x)}\,,
\ean
which we called the reverse Kullback-Leibler distance in the previous Section. Only the second term depends on the reference prior; besides, it does not depend on $q(y)$, but does depend on $p(x)$ that may be arbitrary. We may then choose $\xi$ as to \textit{minimize the average of $D_{KL}(P_F||P_R[\xi])$ over all possible choices of $p$}. Upon such averaging, $p(x)\rightarrow\frac{1}{d}$. Thus we want to find $\xi_{D}=\text{argmin}F[\xi]$
with
\ba\label{eq:mindkl}
F[\xi]&=&\sum_{x,y}\varphi(y|x)\ln\frac{\hat{\xi}(y)}{\xi(x)}\,.
\ea
Interestingly, we have:
\valerio{\begin{result}\label{res4}
The reference prior that minimises \eqref{eq:mindkl} is $\xi_D(x)=\frac{1}{d}$ the uniform prior, for every $\varphi$.
\end{result}
We present the proof in Appendix \ref{app:minKL}.}

In the same spirit, one can study the reference priors that minimize other figures of merit averaged over the possible priors $p(x)$ and $q(y)$. We run some simple numerical checks at $d=2$ for two other figures of merit. For the \textit{Kullback-Leibler distance} $D_{KL}(P_R[\xi]||P_F)$, the steady state is generically not optimal, while the uniform prior seems to be optimal again, even though the dependence on $\xi$ is different from \eqref{eq:mindkl}. For the \textit{guessing probability}, i.e.~the probability that $\textrm{argmax}P_F(x,y)=\textrm{argmax}P_R[\xi](x,y)$, neither the steady state nor the uniform prior are generically optimal.

\section{Final considerations: Do we need a reverse physical process?}\label{sec:conclusion}

The everyday meaning of (ir)reversibility in nature is captured by the perceived ``arrow of time'': if the video of the evolution played backward makes sense, the process is reversible; if it doesn't make sense, it is irreversible.

Science has gone very far in bringing this intuition on quantitative ground. The standard underlying narrative still involves two processes: the one that we observe, and the associated reverse process (not deemed to be strictly impossible, but very unlikely). This reverse process is generically \textit{not} the video played backward: to cite an extreme example, nobody conceives bombs that fly upward to their airplanes while cities are being built from rabble \cite{Earman_1974, BarrettSober}. In the case of controlled protocols in the presence of an unchanging environment, the reverse process is implemented by reversing the protocol. If the environment were to change (in an uncontrolled way, by definition of environment), the connection between the physical process and the associated reverse one becomes thinner.

With our line of research, we are exploring the possibility that the narrative of the reverse process may not be needed at all. In the wording pioneered by Watanabe, irreversibility may be rather \textit{irretrodictability}. So far, this program has found no obstacle, and has even clarified situations that were deemed puzzling in the case of some quantum channels \cite{BS11}. The vistas opened by this approach also allow to expand the scope of fluctuation relations (Section \ref{sec:ft}) and discuss the choice of a reference prior (Section \ref{sec:priors}).

Barring surprises \`a la John Bell, this conflict of narratives won't be discriminated by experiments. Indeed, on the one hand, the retrodiction channels (both classical and quantum) are by construction valid channels: nothing forbids the physical implementation of the corresponding processes, as indeed was done in the experimental verifications of Crooks' theorem \cite{Collin05}. On the other hand, to falsify the retrodictive narrative, one would have to find a reverse process related to its original process in a way that cannot be expressed by (or worse, contradicts) logical reasoning: it is hard to see how such a claim could ever be made. So, one's narrative of choice will depend on the fruitfulness of the intuition, the economy of concepts, the elegance of the formulas... In this paper, we have hinted at the superiority of the retrodictive narrative in all these respects.

\section*{Acknowledgments}

\valerio{We acknowledge the help of Eugene Koh in completing the proof of Result \ref{res4}.} F.B. acknowledges support from the Japan Society for the Promotion of Science (JSPS) KAKENHI, Grants Nos.19H04066 and 20K03746, and from MEXT Quantum Leap Flagship Program (MEXT Q-LEAP), Grant Number JPMXS0120319794. V.S. acknowledges support from the National Research Foundation and the Ministry of Education, Singapore, under the Research Centres of Excellence programme.

Data sharing is not applicable to this article as no new data were created or analyzed in this study.

\begin{appendix}

\section{The reference prior that minimizes the average Kullback-Leibler distance}\label{app:minKL}

\valerio{Consider a generic classical channel $\varphi$ with $d$-dimensional input and output alphabets. Denote the reference prior as $\xi(x)=\frac{1}{d}(1+u_x)$ with $-1\leq u_x\leq d-1$ and $\sum_xu_x=0$. With this parametrisation, $F[\xi]\equiv F(\underline{u})$ with
\ban
F(\underline{u})&=&\sum_{xy}\varphi(y|x)\ln\left(\frac{\sum_{x'}\varphi(y|x')(1+u_{x'})}{1+u_x}\right)\,.
\ean
On the uniform prior ($u_x=0$ for all $x$), this takes the value
\ban
F(\underline{0})&=&\sum_{xy}\varphi(y|x)\ln\left(\sum_{x'}\varphi(y|x')\right)\,.
\ean
Thus, $\Delta\equiv F(\underline{u})-F(\underline{0})$ is equal to
\ban
\Delta&=&\sum_{xy}\varphi(y|x)\ln\left(\frac{\sum_{x'}\varphi(y|x')(1+u_{x'})}{\sum_{x'}\varphi(y|x')}\right)-\sum_x\ln(1+u_x)\\
&\geq& \sum_{xy}\varphi(y|x)\frac{\sum_{x'}\varphi(y|x')\ln(1+u_{x'})}{\sum_{x'}\varphi(y|x')}-\sum_x\ln(1+u_x)\\
&=&\sum_{x,y}\varphi(y|x)\ln(1+u_{x})-\sum_x\ln(1+u_x)\,=\,0
\ean
where the inequality is Jensen's inequality on the first term. Thus we have proved that 
\ba
F(\underline{u})-F(\underline{0})\geq 0\;&&\;\textrm{for all }\underline{u}\,.
\ea

}

\end{appendix}

\section*{References}
\bibliography{refsretro}

\end{document}